\documentclass[%
reprint,
superscriptaddress,
preprintnumbers,
amsmath,amssymb,
aps,
prd,
]{revtex4-2}

\usepackage[utf8]{inputenc}

\usepackage{mathtools}
\usepackage{amsfonts}
\usepackage{mathrsfs}
\usepackage{bm}
\usepackage{bbold}
\usepackage{slashed}
\usepackage{dsfont}
\usepackage{float}
\usepackage{dcolumn}
\usepackage{amssymb,amsmath}

\usepackage{graphicx}
\usepackage{subcaption}
\usepackage{color}
\usepackage{array}

\usepackage{placeins}
\usepackage{booktabs}
\usepackage{caption}

\usepackage{xspace}
\usepackage{hyperref}
\usepackage[nameinlink]{cleveref}
\usepackage{bookmark}
\usepackage{units}

\def\Eq#1{Eq.~\labelcref{#1}}

\def\Fig#1{Fig.~\labelcref{#1}}
\def\Tab#1{Tab.~\labelcref{#1}}
\def\sec#1{Sec.~\labelcref{#1}}

\setkeys{Gin}{width=0.48\textwidth}
\usepackage{booktabs}
\usepackage{multirow}

\newcolumntype{C}{>{$}c<{$}}
\AtBeginDocument{
	\heavyrulewidth=.08em
	\lightrulewidth=.05em
	\cmidrulewidth=.03em
	\belowrulesep=.65ex
	\belowbottomsep=0pt
	\aboverulesep=.4ex
	\abovetopsep=0pt
	\cmidrulesep=\doublerulesep
	\cmidrulekern=.5em
	\defaultaddspace=.5em
}

\graphicspath{{./figures/}}

\usepackage{xifthen}
\usepackage{xcolor}

\newcommand{\gettitle}{Reconstruction of baryon number distributions}

\hypersetup{
	colorlinks,
	linkcolor={red!75!black},
	citecolor={blue!75!black},
	urlcolor={blue!75!black}, 
	pdftitle={\gettitle},
	pdfauthor={},
	pdfkeywords={}
	{} 
	bookmarksopen=true,
	bookmarksopenlevel=2,
	bookmarksnumbered=true
}

\begin{document}
\title{\gettitle}

\author{Chuang Huang}
\affiliation{School of Physics, Dalian University of Technology, Dalian, 116024,
  P.R. China}	

\author{Yang-yang Tan}
\affiliation{School of Physics, Dalian University of Technology, Dalian, 116024,
  P.R. China}	

\author{Rui Wen}
\affiliation{School  of  Nuclear  Science  and  Technology, University  of  Chinese  Academy  of  Sciences,  Beijing,  P.R.China  100049}	

\author{Shi Yin}
\affiliation{School of Physics, Dalian University of Technology, Dalian, 116024,
  P.R. China}	

\author{Wei-jie Fu}
\email{wjfu@dlut.edu.cn}
\affiliation{School of Physics, Dalian University of Technology, Dalian, 116024,
  P.R. China}

	
\begin{abstract}

The maximum entropy method (MEM) and the Gaussian process (GP) regression, which are both well-suited for the treatment of inverse problems, are used to reconstruct net-baryon number distributions based on a finite number of cumulants of the distribution. Baryon number distributions across the chiral phase transition are reconstructed. It is found that with the increase of the order of cumulants, distribution in the long tails, i.e., far away from the central number, would become more and more important. We also reconstruct the distribution function based on the experimentally measured cumulants at the collision energy $\sqrt{s_{NN}}=7.7$ GeV. Given the sizable error of the fourth-order cumulant measured in experiments, the calculation of MEM shows that with the increasing fourth-order cumulant, there is another peak in the distribution function developed in the region of large baryon number. This unnaturalness observed in the reconstructed distribution function might in turn be used to constrain the cumulants measured in experiments.

\end{abstract}

\maketitle
	
\section{Introduction}
\label{sec:int}

In the QCD phase diagram in the plane of temperature and baryon chemical potential, it is conjectured that there is a critical end point (CEP) connecting the first-order phase transition line at high baryon chemical potential and the continuous crossover in the regime of high temperature \cite{Aoki:2006we, Andronic:2017pug}, see e.g., \cite{Stephanov:2007fk, Fischer:2018sdj, Fu:2022gou} for related reviews. Due to the tremendous importance of CEP in the studies of QCD phase structure, that would deepen our understanding of the fundamental properties of strongly correlated QCD matter in extreme conditions, recent years have seen significant progress in the studies of CEP from both the theoretical and experimental sides.

Lattice QCD simulations indicate that there is no CEP in the region of $\mu_B/T\lesssim 2 \sim 3$ \cite{Karsch:2019mbv}, where $\mu_B$ denotes the baryon chemical potential and $T$ the temperature. Reliable computations of lattice QCD at larger $\mu_B$ are hindered by the rapidly increasing severity of the sign problem at finite chemical potential, see e.g., \cite{Borsanyi:2021sxv, Borsanyi:2022soo, Bollweg:2022rps} for more details. The region of reliable calculations can be further extended to that of $\mu_B/T\sim 4$ within the functional continuous field approaches, for instance the functional renormalization group (fRG) \cite{Wetterich:1992yh, Pawlowski:2005xe, Fu:2022gou, Dupuis:2020fhh} and Dyson-Schwinger equations \cite{Fischer:2018sdj}, which should have passed through benchmark tests of lattice QCD in the regime of small $\mu_B$. It is interesting to note that recent first-principles functional approaches have shown a convergent estimate of the location of CEP in the phase diagram at a small region around about $\mu_B \sim 600$ MeV \cite{Fu:2019hdw, Gao:2020fbl, Gunkel:2021oya}.

The QCD critical end point is a second-order phase transition point, that belongs to the $Z(2)$ symmetry, i.e., Ising-like, universality class. Generally speaking, fluctuation measurements, in particular high-order non-Gaussian fluctuations, are sensitive to the critical behavior of the second-order phase transition. Thus, it has been proposed for a long time that one can use the fluctuations of conserved charges, such as the baryon number fluctuations, to search for the CEP \cite{Stephanov:1999zu, Stephanov:2008qz, Stephanov:2011pb}. The relevant experiments have been underway at the Relativistic Heavy Ion Collider (RHIC) over the last decade, and fluctuations of the proton, electric charge and net-kaon numbers have been measured \cite{Adamczyk:2013dal, Adamczyk:2014fia, Luo:2015ewa, Adamczyk:2017wsl, Adam:2019xmk, STAR:2020tga, STAR:2021rls, STAR:2021fge, STAR:2023esa, Pandav:2023lis}. Remarkably, a non-monotonic dependence of the kurtosis of the net-proton distributions, i.e., the fourth-order proton number fluctuations, on the collision energy is observed with $3.1\,\sigma$ significance \cite{STAR:2020tga}. Although this observation can not yet provide evidence for the existence of a CEP in the QCD phase diagram \cite{Braun-Munzinger:2020jbk, Fu:2021oaw}, it is indeed an important progress in this direction.

The critical behavior of a phase transition is encoded in the cumulants of net-proton or net-baryon number distributions, and hence these properties should also be reflected directly in the distribution function itself. In comparison to cumulants, more information is included in the baryon number distributions, rather than a few moments of the distribution function. Moreover, the baryon number distributions can be used in some transport models of heavy-ion collisions, see e.g., \cite{Chen:2022wkj, Chen:2022xpm}. Therefore, it is very interesting and useful to investigate the baryon number distributions directly.

The baryon number distribution can be computed directly through calculation of the canonical partition function, see e.g., \cite{Sun:2018ozp} for more details. The calculation, however, becomes more and more difficult with the increase of the baryon chemical potential, and it is almost inaccessible for the chemical freeze-out baryon chemical potentials in the Beam Energy Scan experiments at RHIC. In this work, we would like to adopt an alternative approach, and use a few order cumulants to reconstruct the baryon number distributions. This task is a typical inverse problem, that is ill-defined. Thanks to the rapid progress in statistics and computing science, a number of methods have been developed to deal with inverse problems. In this work, we choose two of them, the maximum entropy method (MEM) and the Gaussian process regression (GPR) or just abbreviated as Gaussian process (GP), to reconstruct the baryon number distributions from a finite number of cumulants. Eventually, We would like to apply the reconstruction to cumulants which are obtained in both theoretical calculations and experimental measurements.

This paper is organized as follows: In \sec{sec:Thermo} we give a brief introduction about the relation between the baryon number distribution and the generalized baryon number susceptibilities. The approaches of MEM and GP are discussed in \sec{sec:MaxEnt} and \sec{sec:GPR}, respectively. In \sec{sec:NumSetup} numerical setups for these two approaches are given. In  \sec{sec:Num-resu} we use the two methods to reconstruct the baryon number distributions across the chiral phase transition and those at collision energy $\sqrt{s_{NN}}=7.7$ GeV. Finally, conclusions and summary are presented in \sec{sec:conclu}.

\section{Thermodynamics and baryon number fluctuations}
\label{sec:Thermo}

We proceed with a basic relation between fluctuations and the probability distribution for a conserved charge, e.g., the net baryon number $N_{B}$. The $n$-th order central moment of a net baryon number distribution is defined as
\begin{align}
    \langle(\delta N_{B})^{n}\rangle=&\sum_{N_{B}=-\infty}^{\infty}(\delta N_{B})^{n}P(N_{B})\,, \label{eq:cum-def}
\end{align}
with $\delta N_{B}\equiv N_{B}-\langle N_{B} \rangle$, where $P(N_{B})$ denotes the probability distribution of the net baryon number $N_{B}$, and $\langle N_{B} \rangle$ is its mean value. Consequently, if the probability distribution is known, one can readily obtain the relevant cumulants of different orders, cf. e.g. \cite{Morita:2012kt, Morita:2013tu, Sun:2018ozp}.

For a system of grand canonical ensemble in thermal equilibrium, the distribution cumulants of a conserved charge can be directly obtained by differentiating the thermodynamic potential of the system w.r.t. the chemical potential conjugate to the charge. Thus, the cumulants of the baryon number are provided by the generalized susceptibilities related to the baryon chemical potential $\mu_{B}$, which read
\begin{align}
    \chi_{n}^{B}=&\frac{\partial^{n}}{\partial(\mu_{B}/T)^{n}}\frac{p}{T^{4}}\,, \label{eq:chi-def}
\end{align}
with pressure $p$ and temperature $T$, where the temperature is fixed. The pressure is directly related to the thermodynamic potential density, to wit,
\begin{align}
    p=&-\Omega[T,\mu_{B}]\,. \label{}
\end{align}
The pressure as well as the thermodynamic potential of QCD matter for a grand canonical ensemble in thermal equilibrium can be computed from lattice QCD simulations \cite{Borsanyi:2021sxv, Borsanyi:2022soo, Bollweg:2022rps} and functional approaches, such as the fRG \cite{Pawlowski:2005xe, Fu:2019hdw, Fu:2022gou} and Dyson-Schwinger equations \cite{Fischer:2018sdj}. In the fRG approach, the thermodynamic potential is extracted from the difference between the effective action at finite $T$ and $\mu_B$ and that in the vacuum, that is,
\begin{align}
    \Omega[T,\mu_{B}]=&\frac{T}{V}\Big{(}\Gamma[\bar{\Phi}]\Big{|}_{T,\mu_{B}}-\Gamma[\bar{\Phi}]\Big{|}_{T=\mu_{B}=0}\Big{)}\,, \label{eq:potential}
\end{align}
with volume $V$, where $\Gamma$ is the effective action and $\bar{\Phi}$ denotes a collection of different fields on their respective equations of motion, see, e.g., \cite{Fu:2015naa, Fu:2016tey, Fu:2021oaw} for more details. 

For the lowest four orders, the generalized susceptibilities of baryon number $\chi_{n}^{B}$ in \Eq{eq:chi-def} are related to the cumulants in \Eq{eq:cum-def} through equations as follows,
\begin{subequations}\label{eq:chi14}
\begin{align}
    \chi_{1}^{B}&=\frac{1}{VT^{3}}\langle N_{B} \rangle\,,\\[1ex]
    \chi_{2}^{B}&=\frac{1}{VT^{3}}\langle (\delta N_{B})^{2} \rangle\,,\\[1ex]
    \chi_{3}^{B}&=\frac{1}{VT^{3}}\langle (\delta N_{B})^{3} \rangle\,,\\[1ex]
    \chi_{4}^{B}&=\frac{1}{VT^{3}}\Big{(}\langle (\delta N_{B})^{4}\rangle-3\langle(\delta N_{B})^{2}\rangle^{2}\Big{)}\,.
\end{align}
\end{subequations}
It is also convenient to use the cumulants directly related to $\chi_{n}^{B}$, given by
\begin{align}
    \kappa_{n}=&(VT^{3})\chi_{n}^{B}\,, \label{eq:kappa}
\end{align}
which are proportional to the volume $V$, and thus are extensive. In the following, we will use the cumulants calculated in thermodynamics from \Eq{eq:chi-def}, and try to reconstruct the baryon number distributions in \Eq{eq:cum-def} by employing the MEM and GP.

\section{Maximum entropy method}
\label{sec:MaxEnt}

In this section, we present the distribution reconstruction with the maximum entropy strategy. The maximum entropy method embedded in the Bayesian inference has been successfully applied to ill-posed inverse problems for over 30 years \cite{skilling:1991bayesian, Jarrell:1996rrw, Asakawa:2000tr, Burnier:2013nla}. It's straightforward to apply the MEM to the reconstruction of net baryon number distributions, based on several lowest-order cumulants of the distribution.

We start with a rather general discussion. Suppose that we have some known data of $n$-th order cumulants $\kappa=(\kappa_1, \kappa_2, \cdots, \kappa_n, \cdots)$ and also some prior information $I$ (e.g. the distribution is positive definite), the probability of the distribution of $P(N_{B})$ with the previous conditions is denoted as $\mathscr{P}[P|\kappa,I]$, that is also called as the posterior probability. Then we are left with the task to find a proper distribution $P(N_{B})$, such that the resulting posterior probability $\mathscr{P}[P|\kappa,I]$ reaches its maximum value. The Bayesian theorem tells us that one can access this probability from the likelihood probability and the prior probability in the following way:
\begin{align}
    \mathscr{P}[P|\kappa,I]=&\frac{\mathscr{P}[\kappa|P,I]\mathscr{P}[P|I]}{\mathscr{P}[\kappa|I]}\,, \label{eq:BayeFormu}
\end{align}
where $\mathscr{P}[\kappa|P,I]$ refers to the likelihood probability which describes the observed data distributions and $\mathscr{P}[P|I]$ is the so-called prior probability which is a prior estimate of $P(N_{B})$ without any observed data. The denominator on the right side of \Eq{eq:BayeFormu}, $\mathscr{P}[\kappa|I]$, is a normalization constant that can be safely ignored in the inference process in what follows. The observed data distributions are most commonly assumed to be Gaussian-like, i.e., 
\begin{align}
    \mathscr{P}[\kappa|P,I]\propto &\;\mathrm{e}^{-L}\,, \label{eq:likeliProb}
\end{align}
with
\begin{align}
    L=&\frac{1}{2}\sum_{i,j}(\kappa_i-\kappa^{P}_i)C^{-1}_{ij}(\kappa_j-\kappa^{P}_j)\,, \label{eq:LC}
\end{align}
where $\kappa_i$ indicates an average of the $i$-th order cumulant from the observed data and $\kappa^{P}_i$ with a superscript $P$ that calculated from the distribution $P(N_{B})$. $C_{ij}$ is the covariance matrix for the observed data. If only the likelihood probability in \Eq{eq:likeliProb} is maximized, it would be equivalent to the $\chi^2$-fitting. Unfortunately, for this ill-posed problem, the number of data of cumulants in order of $\mathcal{O}(1)$, is much smaller than that of the baryon number distributions $P(N_{B})$ in order of $\mathcal{O}(100)$ that we would like to reconstruct, which might result in a plethora of spurious solutions for the $\chi^2$-fitting. The prior probability then plays the role as a penalty term. In the context of MEM the prior probability reads
\begin{align}
    \mathscr{P}[P|I]\propto &\;\mathrm{e}^{\alpha S(P)}\,, \label{eq:priorProb}
\end{align}
where $S$ is the entropy of the baryon number distribution $P(N_{B})$. For a discrete probability distribution, we adopt the Shannon entropy as follows 
\begin{align}
    S=&-\sum_{i=-\infty}^{\infty}P_i\ln(P_i)\,, \label{eq:ShannEntro}
\end{align}
where we have used the shorthand notation $P_i\equiv P(N_{B})$. The hyperparameter $\alpha$ in \Eq{eq:priorProb} is used to control the relative weight between the likelihood probability and the prior probability. Finally, we are led to the posterior probability that reads
\begin{align}
    \mathscr{P}[P|\kappa,I]\propto &\;\mathrm{e}^{Q(P)}\,, \qquad Q(P)=\alpha S-L\,, \label{eq:QP}
\end{align}
and thus, one is required to maximize the function $Q(P)$. Since $Q(P)$ is a concave functional of $P$, one is able to use, e.g., convex optimization techniques, to maximize it.

If the errors of cumulants calculated from the generalized susceptibilities of the baryon number in \Eq{eq:chi-def} are neglected, that is, the covariance matrix in \Eq{eq:LC} approaching the limit $C\to 0$, the likelihood probability, i.e., the $L$ part in \Eq{eq:QP}, can be reduced to a few constraints, which read
\begin{align}
    \kappa_n=&\kappa_n^{P}=\sum_{i}K_{ni}P_i\,, \label{eq:kappaK}
\end{align}
where $K_{ni}$ indicates the kernel of cumulant $\kappa_n$. Specifically, if $K_{n=0i}=1$ is chosen, \Eq{eq:kappaK} leaves us with $\kappa_0=1$, that is the normalization for the distribution function. Taking into account the constraints discussed above, the entropy in \Eq{eq:ShannEntro} is extended to the equation as follows
\begin{align}
    S=&-\sum_{i=-\infty}^{\infty}P_i\ln(P_i)+\sum_{n=0}^{n_{\mathrm{trunc}}}\lambda_n\left(\sum_{i=-\infty}^{\infty} K_{ni}P_i-\kappa_n\right)\,, \label{eq:ShannEntro2}
\end{align}
where $\lambda_n$ is the Lagrange multiplier for each constraint, and $n_{\mathrm{trunc}}$ stands for the maximal order of cumulants used to reconstruct the distribution. Maximizing \Eq{eq:ShannEntro2} with respect to the probability distribution $P_i$ and the multipliers $\lambda_n$, one is led to an analytic baryon number distribution which reads
\begin{align}
   P_i=&\exp\left(\sum_{n=1}^{n_{\mathrm{trunc}}}\lambda_n K_{ni}+\lambda_0-1\right)\,, \label{eq:MaxEntP}
\end{align}
where the Lagrange multipliers are determined by the constraints in \Eq{eq:kappaK}.

\section{Gaussian process regression}
\label{sec:GPR}

The Gaussian process regression is a powerful method to infer the probability function from a finite number of observations, cf. reviews \cite{kanagawa2018gaussian, liu2020gaussian} and textbook \cite{williams2006gaussian}. Recently, it has been employed in the study of parton distribution functions from lattice QCD \cite{Alexandrou:2020tqq} and the reconstruction of the QCD spectral functions from fRG \cite{Horak:2021syv, Horak:2023xfb}. In this section, the discrete GP, which is suitably applied to the net baryon number distribution, is introduced.

Generally, the probability distribution of the baryon number distribution can be described by a discrete Gaussian process, 
\begin{align}
    P(N_{B})\sim \mathcal{GP}\Big(\mu(N_{B}),C(N_{B},N^{\prime}_{B})\Big)\,,\label{eq:GP}
\end{align}
with the mean function $\mu(N_{B})$ and the covariance function or covariance matrix $C(N_{B},N^{\prime}_{B})$, that is symmetric, i.e., $C(N_{B},N^{\prime}_{B})=C(N^{\prime}_{B},N_{B})$. For the GP, the probability of the function is described by a multivariate Gaussian or normal distribution, which includes a finite set of sample points $\{N_{B_{1}},N_{B_{2}},...,N_{B_{n}}\}$, that is,
\begin{align}
\left(\begin{array}{c}
P(N_{B_{1}}) \\
\vdots \\
P(N_{B_{n}})
\end{array}\right) & \sim \mathcal{N}\left(\left(\begin{array}{c}
\mu(N_{B_{1}}) \\
\vdots \\
\mu(N_{B_{n}})
\end{array}\right),\right.\nonumber\\[2ex]
&\left.\left(\begin{array}{ccc}
C(N_{B_{1}}, N_{B_{1}}) & \ldots & C(N_{B_{1}}, N_{B_{n}}) \\
\vdots & \ddots & \vdots \\
C(N_{B_{n}}, N_{B_{1}}) & \ldots & C(N_{B_{n}}, N_{B_{n}})
\end{array}\right)\right) \,, \label{eq:GP-def}
\end{align}
where $\mathcal{N}$ stands for a $n$-dimensional normal distribution. In particular, this approach is constructed from a complete family of functions, so that there is no restriction on the functional basis or no artificial truncation.

For the reconstruction of baryon number distributions, it is necessary to add a set of observations, here denoted by $O_i$, to \Eq{eq:GP-def}, which yields
\begin{align}
    \left(\begin{array}{c}
P(N_{B}) \\
\bm{O}
\end{array}\right) &\sim \mathcal{N}\Bigg(\left(\begin{array}{c}
\mu(N_{B}) \\
\bm{\mu}_{O}
\end{array}\right),\nonumber\\[2ex]
&\left(\begin{array}{cc}
C\left(N_{B}, N_{B}^{\prime}\right) & \bm{C}^\top(N_{B}) \\
\bm{C}\left(N^{\prime}_{B}\right) & \bm{C}+\sigma_n^2 \cdot \bm{1}
\end{array}\right)\Bigg)\,,\label{eq:GP-addobs}
\end{align}
with 
\begin{align}
    {\bm{\mu}_{O}}_i&\equiv\mu(O_i)\,,\qquad \bm{C}_i(N_{B})\equiv C(O_i,N_{B})\,,\label{eq:muO}\\[2ex]
    \bm{C}_{ij}&\equiv C(O_i,O_j)\,. \label{eq:Cij}
\end{align}
where $\sigma_n^2$ denotes the pointwise variance due to the measurement noise in the observations. Note that the observation can be either the baryon number distribution $P(N_{B})$ itself at some values of $N_{B}$, or the cumulants of the distribution as shown in \Eq{eq:kappaK}. Thus, given the posterior information on baryon number distributions provided by the observations $O_i$, one is led to the conditional probability of baryon number distributions \cite{williams2006gaussian}, as follows
\begin{align}
  P(N_{B}) \big | \bm{O} \sim&\mathcal{N}\Big(\bar\mu(N_{B}),\bar C(N_{B}, N_{B}^{\prime})\Big)\,,\label{eq:GP-condi}
\end{align}
with 
\begin{align}
  &\bar\mu(N_{B}) \nonumber\\[2ex]
  \equiv &\mu(N_{B})+\bm{C}^{\top}(N_{B})\Big(\bm{C}+\sigma_n^2 \cdot \bm{1}\Big)^{-1}(\bm{O}-\bm{\mu}_{O})\,,\label{eq:barmu}\\[2ex] 
  &\bar C(N_{B}, N_{B}^{\prime})\nonumber\\[2ex]
  \equiv&C(N_{B}, N_{B}^{\prime})-\bm{C}^{\top}(N_{B})\Big(\bm{C}+\sigma_n^2 \cdot \bm{1}\Big)^{-1} \bm{C}(N_{B}^{\prime}) \,,\label{eq:barC}
\end{align}
which follow from the standard results of multivariate normal distribution. On the other hand, the probability of the observed data also obeys the multivariate normal distribution, which reads 
\begin{align}
  \bm{O}_i \sim &\mathcal{N}\Big({\bm{\mu}_{O}}_i, \bm{C}_{ij}\Big)\,,\label{eq:N-obs}
\end{align}
with the mean value in \Eq{eq:muO} and the covariance in \Eq{eq:Cij} given by
\begin{align}
  {\bm{\mu}_{O}}_i = &\sum_{N_{B}} K_{i}(N_{B})\,\mu(N_{B})\,,\label{eq:muO-K}\\[2ex]
  \bm{C}_{ij}=&\sum_{N_{B},N'_{B}} K_{i}(N_{B}) C(N_{B}, N_{B}^{\prime})K_{j}(N'_{B})\,,\label{eq:Cij-K}
\end{align}
where $K_{i}(N_{B})$ corresponds to the kernel in \Eq{eq:kappaK} if the cumulants are observed. In the same way, one also has 
\begin{align}
  \bm{C}_i(N_{B})=&\sum_{N_{B}^{\prime}} K_{i}(N'_{B}) C(N'_{B}, N_{B})\,.\label{eq:CiNB-K}
\end{align}
Inserting \Eq{eq:muO-K}, \Eq{eq:Cij-K}, and \Eq{eq:CiNB-K} into \Eq{eq:barmu} and \Eq{eq:barC}, one arrives at
\begin{align}
  \bar\mu(N_{B})= &\mu(N_{B})+\sum_{i,j}\sum_{N_{B}^{\prime}} K_{i}(N'_{B}) C(N'_{B}, N_{B})\nonumber\\[2ex]
  &\hspace{1.5cm}\times\Big(\bm{C}+\sigma_n^2 \cdot \bm{1}\Big)^{-1}_{ij}(\bm{O}_j-{\bm{\mu}_{O}}_j)\,,\label{eq:barmu2}\\[2ex] 
  \bar C(N_{B}, N_{B}^{\prime})=&C(N_{B}, N_{B}^{\prime})-\sum_{i,j}\sum_{N''_{B}N'''_{B}} K_{i}(N''_{B}) C(N''_{B}, N_{B})\nonumber\\[2ex]
  &\hspace{0.5cm}\times\Big(\bm{C}+\sigma_n^2 \cdot \bm{1}\Big)^{-1}_{ij} K_{j}(N'''_{B}) C(N'''_{B}, N'_{B}) \,,\label{eq:barC2}
\end{align}
where the summation over the subscripts $i,j$ applies for all observed data.

%
\begin{figure}[t]
\includegraphics[width=0.45\textwidth]{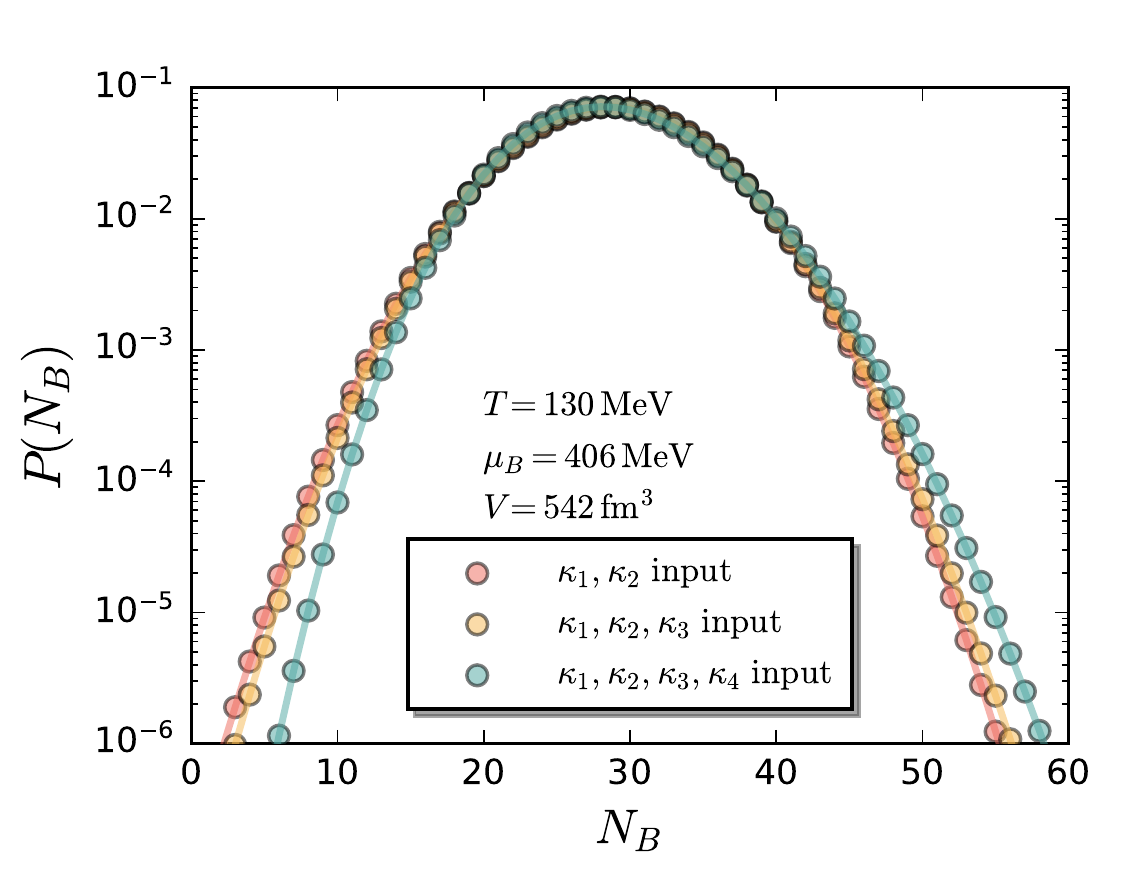}
\caption{Baryon number distributions reconstructed from the cumulants obtained in fRG \cite{Fu:2021oaw} with $T=130$ MeV, $\mu_B=406$ MeV, and $V=542\,\mathrm{fm}^3$ by means of the MEM. Three different inputs are employed, corresponding to the input of only $\kappa_1$ and $\kappa_2$, that of $\kappa_1$ through $\kappa_3$, and that of $\kappa_1$ through $\kappa_4$, respectively.}\label{fig:rho-MaxEnt}
\end{figure}
%

In this work, we adopt the commonly used kernel function called the radial basis function (RBF), which reads
\begin{align}
  C(N_{B}, N'_{B})=&\sigma_c^2 \exp \bigg(-\frac{(N_{B}-N'_{B})^2}{2 l^2}\bigg)\,,\label{eq:RBF}
\end{align}
where two hyperparameters denoted by $\bm{\alpha}=\{\sigma_c,l\}$ have been introduced, which are determined by minimizing the negative logarithm likelihood (NLL) probability in \Eq{eq:N-obs}, to wit,
\begin{align}
  -\log\mathscr{P}[\bm{O};\bm{\alpha}]=&\frac{1}{2}(\bm{O}-\bm{\mu}_{O})^{\top}\Big(\bm{C}_{\bm{\alpha}}+\sigma_n^2 \cdot \bm{1}\Big)^{-1}(\bm{O}-\bm{\mu}_{O})\nonumber\\[2ex]
  &+\frac{1}{2}\log\det\Big(\bm{C}_{\bm{\alpha}}+\sigma_n^2 \cdot \bm{1}\Big)+\frac{N}{2}\log(2\pi)\,,\label{eq:NLL}
\end{align}
where the explicit dependence on the hyperparameters is shown, and $N$ is the number of the observations.

\section{Numerical setup}
\label{sec:NumSetup}

%
\begin{table*}[t]
  \begin{center}
  \begin{tabular}{cccccccccc}
    \hline\hline & & & & & & & & &  \\[-2ex]   
    & $\kappa_1$ & $\kappa_2$ & $\kappa_3$ & $\kappa_4$& $\lambda_0$ & $\lambda_1$ & $\lambda_2$ & $\lambda_3$ &$\lambda_4$\\[1ex]
    \hline & & & & & & & & & \\[-2ex]
    fRG input & 28.74 & 31.46 & 37.54 & 54.60 & --- & ---&--- & --- & ---   \\[1ex]
    MEM ($\kappa_1$,$\kappa_2$ input) & 28.74 & 31.46 & $\sim 0^*$ & $\sim 0^*$ & $-1.643$ & $\sim 0$ &$ -1.589 \times 10^{-2}$ &---& ---\\[1ex]
    MEM ($\kappa_1$,$\kappa_2$,$\kappa_3$ input) & 28.74 & 31.46 & 37.54 & ${1.099\times 10^4 }^*$  & $-1.526$& $-4.033\times 10^{-3}$ &$-1.595\times 10^{-2}$ &$4.276\times 10^{-5} $&---\\[1ex]
    MEM ($\kappa_1$,$\kappa_2$,$\kappa_3$,$\kappa_4$ input) & 28.74 & 31.46 & 37.54 & 54.60& $-1.086$&$-1.978\times 10^{-2}$&$-1.567\times 10^{-2}$& $2.147\times 10^{-4}$&  $-3.182\times 10^{-6}$\\[1ex]
    GP ($\kappa_1$,$\kappa_2$,$\kappa_3$,$\kappa_4$ input) & 28.74 & 31.46 & 37.33 & 54.75&  ---& ---& ---&  ---&  ---\\[1ex]
   \hline\hline
  \end{tabular}
  \caption{Reconstruction of the baryon number distributions from several low-order cumulants obtained in fRG \cite{Fu:2021oaw} within the MEM and GP. Numbers with a star are calculated with the reconstructed distribution. Values of the Lagrange multiplies in \Eq{eq:MaxEntP} in the MEM are presented. See text for details.}
  \label{tab:setup}
  \end{center}\vspace{-0.5cm}
\end{table*}
%

%
\begin{figure}[t]
\includegraphics[width=0.45\textwidth]{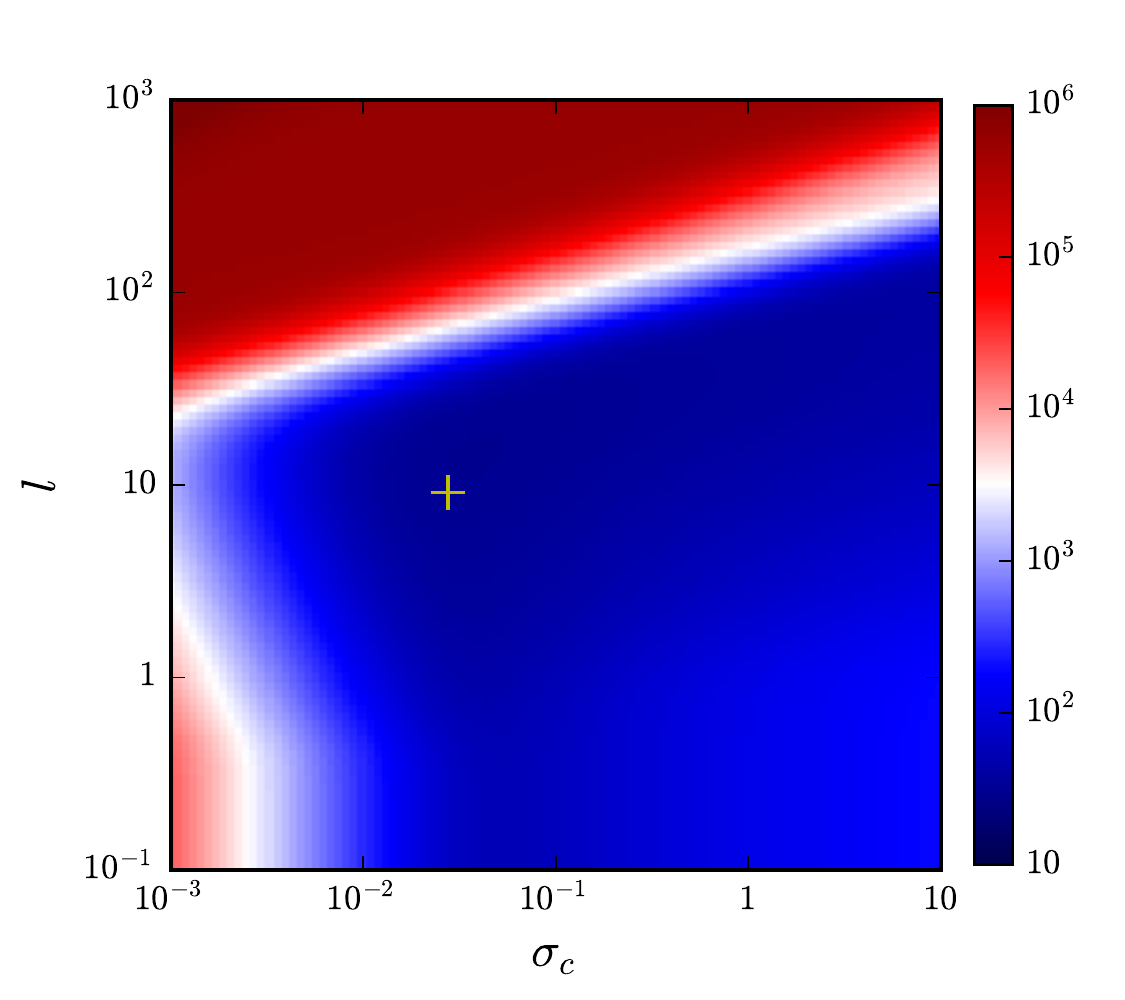}
\caption{Heatmap of the NLL probability as a function of the hyperparameters $\sigma_{c}$ and $l$, where the minimum is labeled with a cross.}\label{fig:NLL}
\end{figure}
%

%
\begin{figure}[t]
\includegraphics[width=0.45\textwidth]{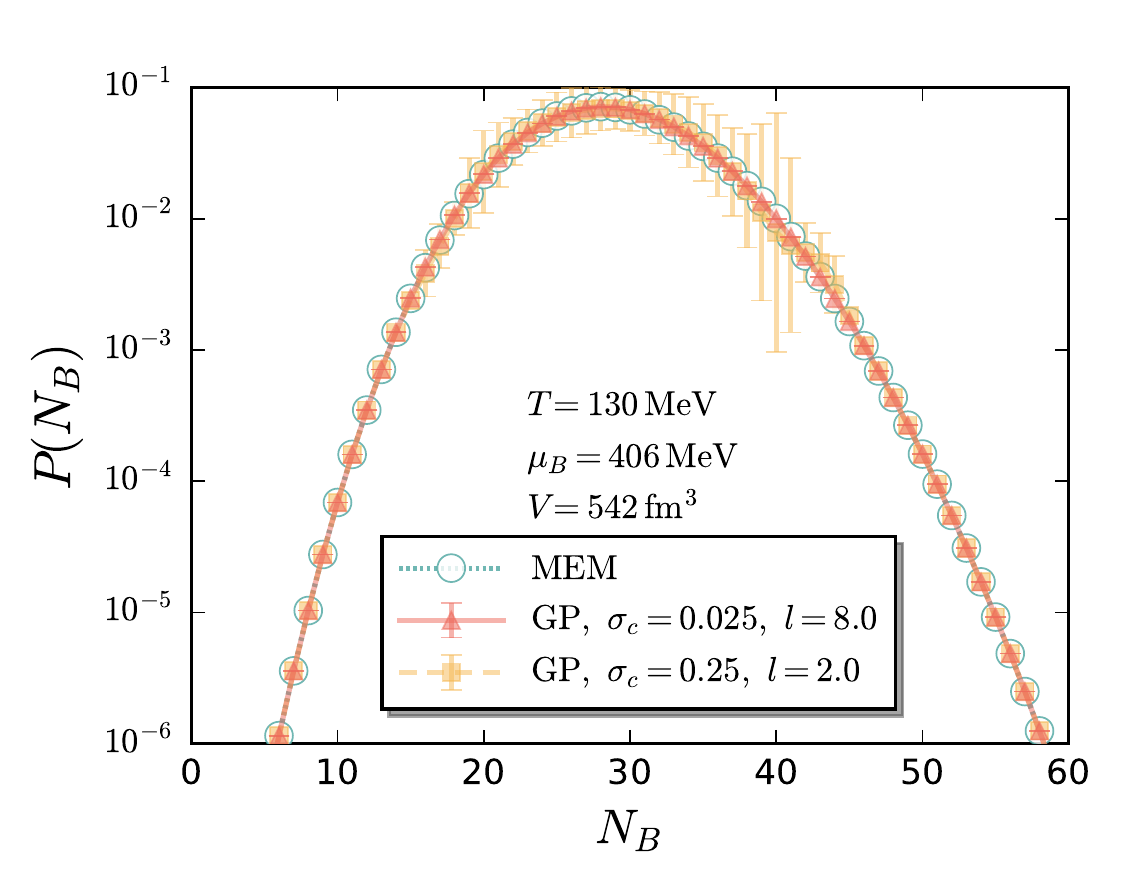}
\caption{Baryon number distributions reconstructed from the first four-order cumulants obtained in fRG \cite{Fu:2021oaw} with $T=130$ MeV, $\mu_B=406$ MeV, and $V=542\,\mathrm{fm}^3$ by means of the GP. The NLL function is minimized  with the hyperparameters $\sigma_{c}=0.025$ and $l=8.0$. The distribution with $\sigma_{c}=0.25$ and $l=2.0$ and that reconstructed through the MEM are presented for comparison.}\label{fig:rho-GP}
\end{figure}
%

In this section, we use a concrete example to illustrate how the baryon number distribution is reconstructed based on a few first low-order cumulants, by means of the MEM and the GP. Here, we use the first four-order cumulants of baryon number distributions obtained in fRG through \Eq{eq:chi-def} at temperature $T=130$ MeV and baryon chemical potential $\mu_B=406$ MeV \cite{Fu:2021oaw}. The volume in \Eq{eq:kappa} is chosen to be $V=542\,\mathrm{fm}^3$, and the resulting values for the first four-order cumulants are presented in the line of fRG input in \Tab{tab:setup}.

For the MEM, we employ three different inputs to reconstruct the baryon number distribution: One is to use just the first two cumulants $\kappa_1$, $\kappa_2$ as inputs; the next one adds $\kappa_3$ into the inputs; the last one adds $\kappa_4$ further. The resulting baryon number distributions for the three different inputs are presented in \Fig{fig:rho-MaxEnt}. One can see clearly that the difference arises in the tails of the distribution, i.e., the region far away from the mean value, when more information of high-order cumulants is encoded. In \Tab{tab:setup}, we also show $\kappa_3$ and $\kappa_4$ of the baryon number distribution reconstructed from $\kappa_1$ and $\kappa_2$, indicated with a star, and $\kappa_4$ from $\kappa_1$, $\kappa_2$ and $\kappa_3$. One observes that the baryon number distribution reconstructed from $\kappa_1$ and $\kappa_2$ within the MEM is consistent with a discretized Gaussian distribution, that is, 
\begin{align}
  P(N_B)\sim&\exp\left({-\frac{(N_B-\kappa_ 1)^2}{2\kappa_2}}\right)\,,\label{}
\end{align}
and the non-Gaussian cumulants $\kappa_3$ and $\kappa_4$ are very close to zero. The baryon number distribution reconstructed with $\kappa_1$, $\kappa_2$ and $\kappa_3$, however, gives rise to a very large nonvanishing $\kappa_4$ as shown in \Tab{tab:setup}. More detailed studies indicate that it is an artifact, that always occurs when cumulants of odd number are used to reconstruct the distribution with the MEM. Moreover, values of the Lagrange multipliers in \Eq{eq:MaxEntP} for each reconstruction are presented in \Tab{tab:setup}.

As for the reconstruction of baryon number distributions by the GP, it is necessary to provide more observations besides the cumulants. For instance, it is natural to assume that the baryon number distribution is nonnegative, i.e., $P(N_{B})\geq 0$ for each value of $N_{B}$; however, this property sometime is violated in the reconstructed distribution by the GP, if only the cumulants are used.  Note that it is difficult to implement the positivity of functions directly in the GP. Alternatively, in this work we provide boundary conditions of the distribution further for the GP in order to avoid the violation of positivity indirectly, say,
\begin{align}
  P(N_B)\Big|_{\mathrm{GP}}=&P(N_B)\Big|_{\mathrm{MEM}}\quad\mathrm{for}\quad P(N_B)\leq 10^{-10}\,.\label{}
\end{align}
The equation above indicates that the distribution in the GP is chosen to be identical to that obtained in MEM, when the baryon number is far away from the mean value.

For the reconstruction of the GP, we use the same first four-order cumulants obtained in fRG as shown in \Tab{tab:setup}. It is found, however, that the third and fourth cumulants calculated from the reconstructed baryon distribution deviate slightly from the input values, as shown in the last line of \Tab{tab:setup}, which arises from numerical errors in the reconstruction of GP. The hyperparameters $\bm{\alpha}=\{\sigma_c,l\}$ in the kernel function in \Eq{eq:RBF} is determined by minimizing the NLL function in \Eq{eq:NLL}. In \Fig{fig:NLL} the NLL probability is depicted in the plane of $l$ and $\sigma_c$, and the location of the minimum is labeled with a cross, that corresponds to $\sigma_{c}=0.025$ and $l=8.0$. Inserting the relevant values of the hyperparameters into \Eq{eq:barmu2} and \Eq{eq:barC2}, one is able to obtain the central value and the variance of $P(N_B)$ for each value of $N_B$, respectively, which are depicted in \Fig{fig:rho-GP}. In order to show explicitly the effects of hyperparameters, we modify their values a bit, say, $\sigma_{c}=0.25$ and $l=2.0$, and present the relevant baryon distribution in \Fig{fig:rho-GP} for comparison. One observes that the central values are almost not altered, but the variance increases significantly in comparison to the optimized case, where the variance is nearly invisible. In \Fig{fig:rho-GP} the distribution obtained from MEM is also presented, and one can see the baryon number distributions reconstructed with these two different approaches are in good agreement with each other.

\section{Numerical results}
\label{sec:Num-resu}

%
\begin{figure}[t]
\includegraphics[width=0.45\textwidth]{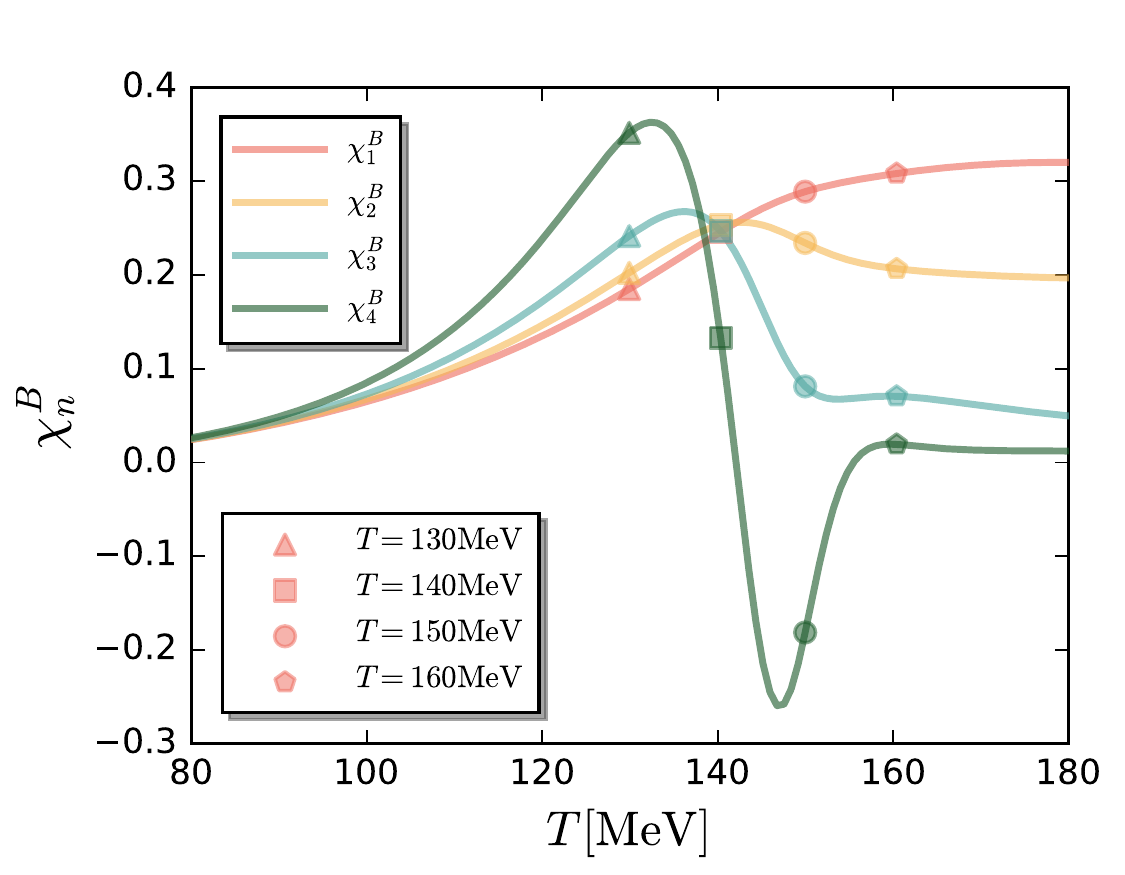}
\caption{Baryon number susceptibilities of the first four orders as functions of the temperature $T$ with the baryon chemical potential $\mu_{B}=406$ MeV, obtained in the fRG computations \cite{Fu:2021oaw}. Susceptibilities at four values of temperature are employed to reconstruct the baryon number distributions in \Fig{fig:rho-T}.}\label{fig:chi-T}
\end{figure}
%

%
\begin{figure}[t]
\includegraphics[width=0.45\textwidth]{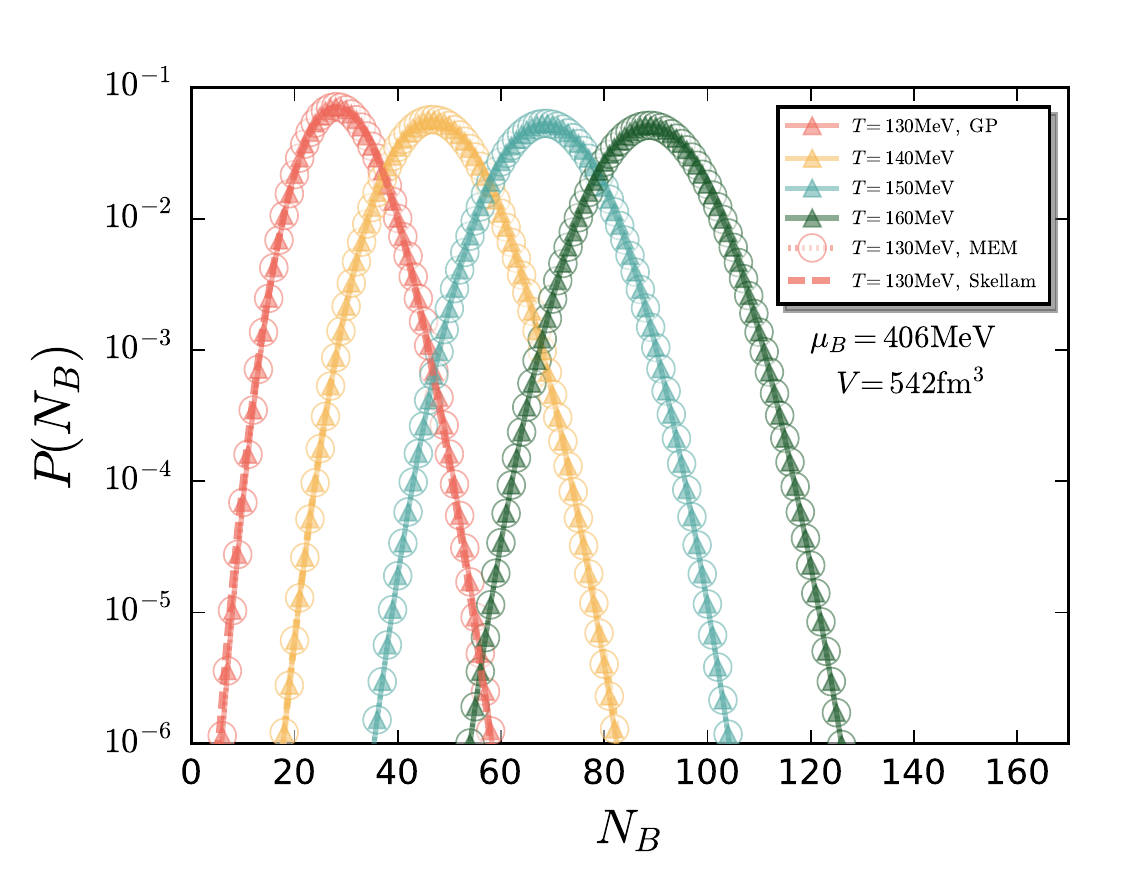}
\caption{Baryon number distributions reconstructed from the cumulants in \Fig{fig:chi-T} at four values of temperature with $\mu_{B}=406$ MeV, where the volume is chosen to be $V=542\,\mathrm{fm}^{3}$. Results of both the MEM and GP are presented. The errors for the distributions that can be computed from the GP are not shown, since they are too small to be visible. The Skellam distribution at $T=130$ MeV is also shown for comparison.}\label{fig:rho-T}
\end{figure}
%

%
\begin{figure*}[t]
\includegraphics[width=0.8\textwidth]{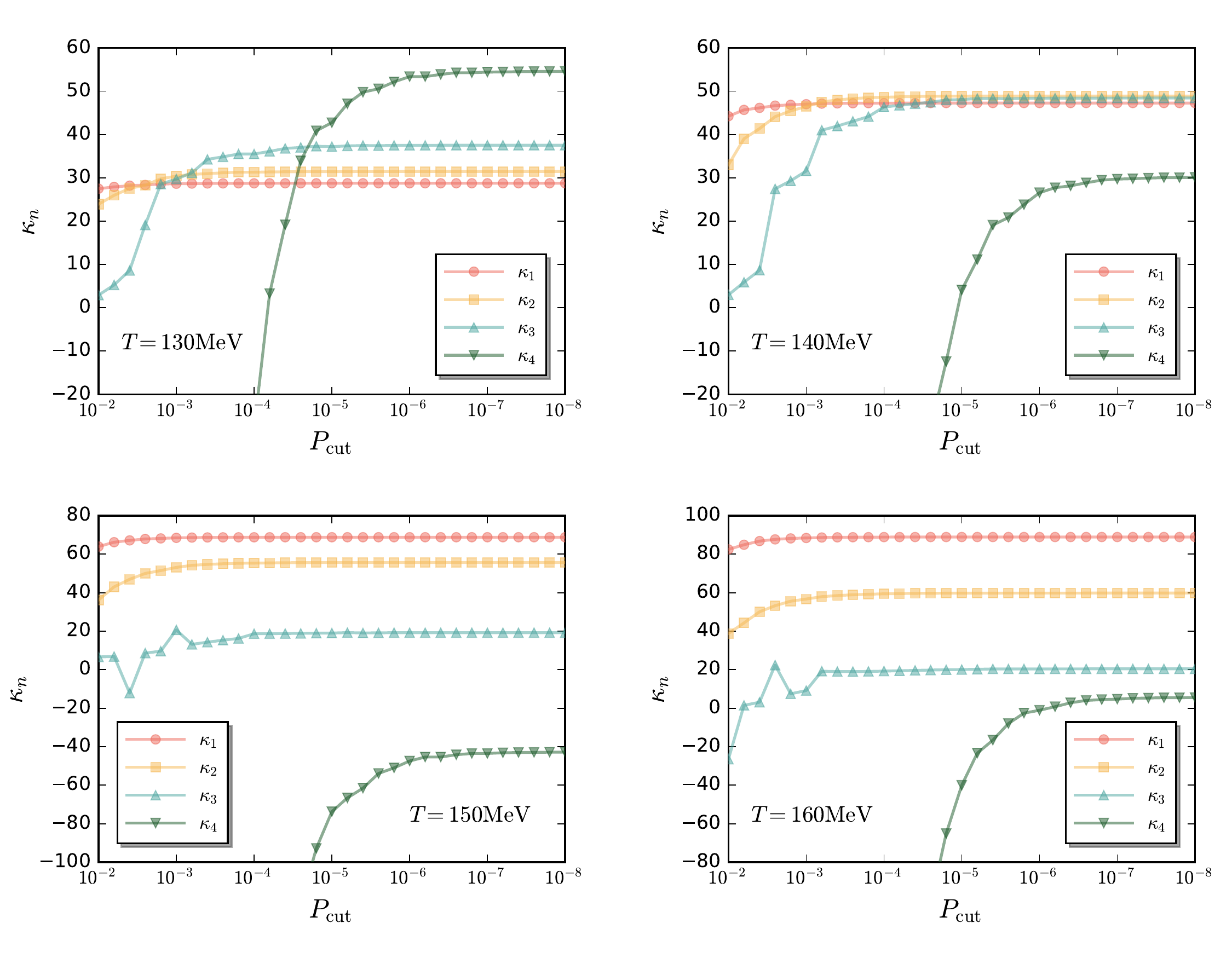}
\caption{Cumulants of the first four orders as functions of the cut $P_{\mathrm{cut}}$ in \Eq{eq:Pcut} at four values of temperature. The reconstructed baryon number distributions in \Fig{fig:rho-T} are used.}\label{fig:Pcut-kappa-T}
\end{figure*}
%

%
\begin{figure*}[t]
\includegraphics[width=0.9\textwidth]{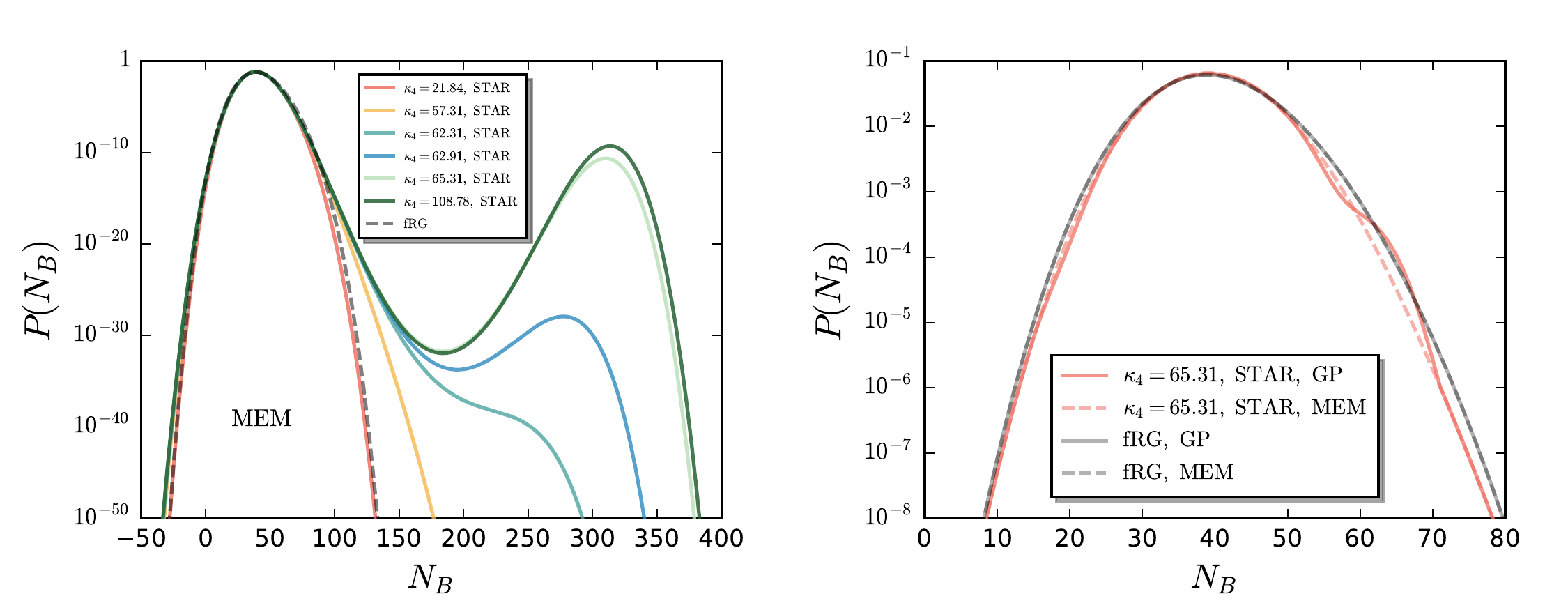}
\caption{Left panel: Baryon number distributions reconstructed within the MEM, where the solid and dashed lines denote the results with the STAR and fRG inputs as shown in \Tab{tab:fRG&Exp-kappa}, respectively. Here, $\kappa_1$, $\kappa_2$, $\kappa_3$ for the STAR data are chosen to be their central values, and $\kappa_4$ is varied around its central value but within its error. Right panel: Comparison between the baryon number distribution reconstructed with the GP and that with the MEM for both the STAR and fRG inputs.}\label{fig:rho-com}
\end{figure*}
%

\subsection{Baryon number distributions across the chiral phase transition}
\label{subsec:Bardistri-transi}

In this section, we use the MEM and GP to reconstruct baryon number distributions across the chiral phase transition, or more exactly, the chiral crossover, with the increase of the temperature at a fixed value of the baryon chemical potential. The employed baryon number susceptibilities of the first four orders are shown in \Fig{fig:chi-T}, which are calculated within the fRG approach, cf. \cite{Fu:2021oaw} for more details. Here we choose the baryon chemical potential $\mu_{B}=406$ MeV, that is around the value of the chemical freeze-out  $\mu_{B}$ with collision energy $\sqrt{s_{NN}}=7.7$ GeV at RHIC. Then, the relevant pseudo-critical temperature of the chiral crossover is around $T_{\mathrm{pc}}\sim 137$ MeV. Four slices of temperature indicated by different symbols as shown in \Fig{fig:chi-T} are chosen. One is smaller than $T_{\mathrm{pc}}$ a bit $T=130$ MeV, and the three others are in the chiral symmetric phase. 

The baryon number distributions corresponding to the four adopted temperatures in \Fig{fig:chi-T} are shown in \Fig{fig:rho-T}. Both the MEM and the GP are used to reconstruct the distributions, whose results are denoted by the triangles and circles, respectively. Once more, one can see that these two methods give rise to consistent distributions. For the case of $T=130$ MeV, i.e., in the hadronic phase and near the chiral phase transition, it is intriguing to compare the reconstructed baryon number distribution with the hadronic Skellam distribution, cf. e.g., \cite{Braun-Munzinger:2011shf}. The Skellam distribution is completely determined by the first two order cumulants, viz.,
\begin{align}
  P(N_B)=&\left(\frac{b}{\bar{b}}\right)^{N_B / 2} I_{N_B}\Big(2 \sqrt{b \bar{b}}\Big)\exp\big[-(b+\bar{b})\big]\,,\label{eq:P-Skellam}
\end{align}
with $b=\kappa_{2}+\kappa_{1}$ and $\bar{b}=\kappa_{2}-\kappa_{1}$ being the mean number of baryons and anti-baryons, respectively. $I_{N_B}$ in \Eq{eq:P-Skellam} is the modified Bessel function of the first kind. The Skellam distribution of $T=130$ MeV is shown in \Fig{fig:rho-T} in the red dashed line. It is found that the reconstructed distribution is consistent with the Skellam distribution except that there is a slight deviation in the left tail of the distribution, since the temperature is in the proximity of $T_{\mathrm{pc}}$. 

If \Eq{eq:cum-def} is modified as such
\begin{align}
    \langle(\delta N_{B})^{n}\rangle=&\sum_{P(N_{B})\geq P_{\mathrm{cut}}}(\delta N_{B})^{n}P(N_{B})\,, \label{eq:Pcut}
\end{align}
that is, a cut for the summation of $N_{B}$ is implemented, one is able to investigate the dependence of the cumulants on the cut $P_{\mathrm{cut}}$, which would reveal some important information, such as how important the tail of the distribution is for different order cumulants. In \Fig{fig:Pcut-kappa-T}, the cumulants of the first four orders are shown as functions of the cut $P_{\mathrm{cut}}$ for the four values of temperature, based on the reconstructed baryon number distributions in \Fig{fig:rho-T}. Since the MEM and the GP have almost the same distributions, we do not distinguish them here. It is clearly shown in \Fig{fig:Pcut-kappa-T} that with the increase of the order of cumulants, the saturation $P_{\mathrm{cut}}$, i.e., the $P_{\mathrm{cut}}$ where the curve of $\kappa_n$ becomes a horizontal line, decreases remarkably. For example, $\kappa_1$ saturates at $P_{\mathrm{cut}}\sim10^{-2}$, $\kappa_2$ $\sim10^{-3}$, $\kappa_3$ $\sim10^{-4}$, $\kappa_4$ $\sim10^{-6}$. This implies that in the experimental measurements of high-order cumulants of the net proton number and other conserved charges, the statistics in the long tails of the distribution would play a pivotal role. Higher order is concerned, longer tail should be taken with care.

\subsection{Baryon number distributions at collision energy $\sqrt{s_{NN}}=7.7$ GeV}
\label{subsec:Bardistri-s7}

%
\begin{table*}[t]
  \begin{center}
 \begin{tabular}{ccccc}
    \hline\hline & & & &  \\[-2ex]   
     & $\kappa_{1}$ & $\kappa_{2}$ & $\kappa_{3}$ & $\kappa_{4}$ \\[1ex]
    \hline & & & &  \\[-2ex]
    STAR\hspace{0.5cm}  & $39.4215{\pm 	0.0200}$ & $36.9630{\pm 0.2142}$ & $29.5181{\pm 2.8394}$ & $65.3126{\pm 43.4652}$  \\[1ex]
    fRG\hspace{0.5cm} & 39.4662 & 42.3560 & 47.6895 & 57.8462  \\[1ex]
    \hline\hline
  \end{tabular}
  \caption{Cumulants of the first four orders of the net-proton distributions for central (0-5\%) Au+Au collisions at the collision energy $\sqrt{s_{NN}}=7.7$ GeV measured by the STAR collaboration \cite{STAR:2020tga}, as well as the theoretical results for the net-baryon number calculated in fRG \cite{Fu:2021oaw}. Note that only the statistical errors of the experimental data are shown here, and we do not include the errors for the fRG results. In the fRG calculations, the chemical freeze-out baryon chemical potential and temperature are ${\mu_B}_{_{\mathrm{CF}}}=406$ MeV and $T_{_{\mathrm{CF}}}=136.2$ MeV, respectively, at $\sqrt{s_{NN}}=7.7$ GeV. The volume is chosen to be $V=542\,\mathrm{fm^3}$ in order to have the same $\kappa_1$ for the experiment and theory.} 
  \label{tab:fRG&Exp-kappa}
  \end{center}\vspace{-0.5cm}
\end{table*}
%

%
\begin{figure}[t]
\includegraphics[width=0.45\textwidth]{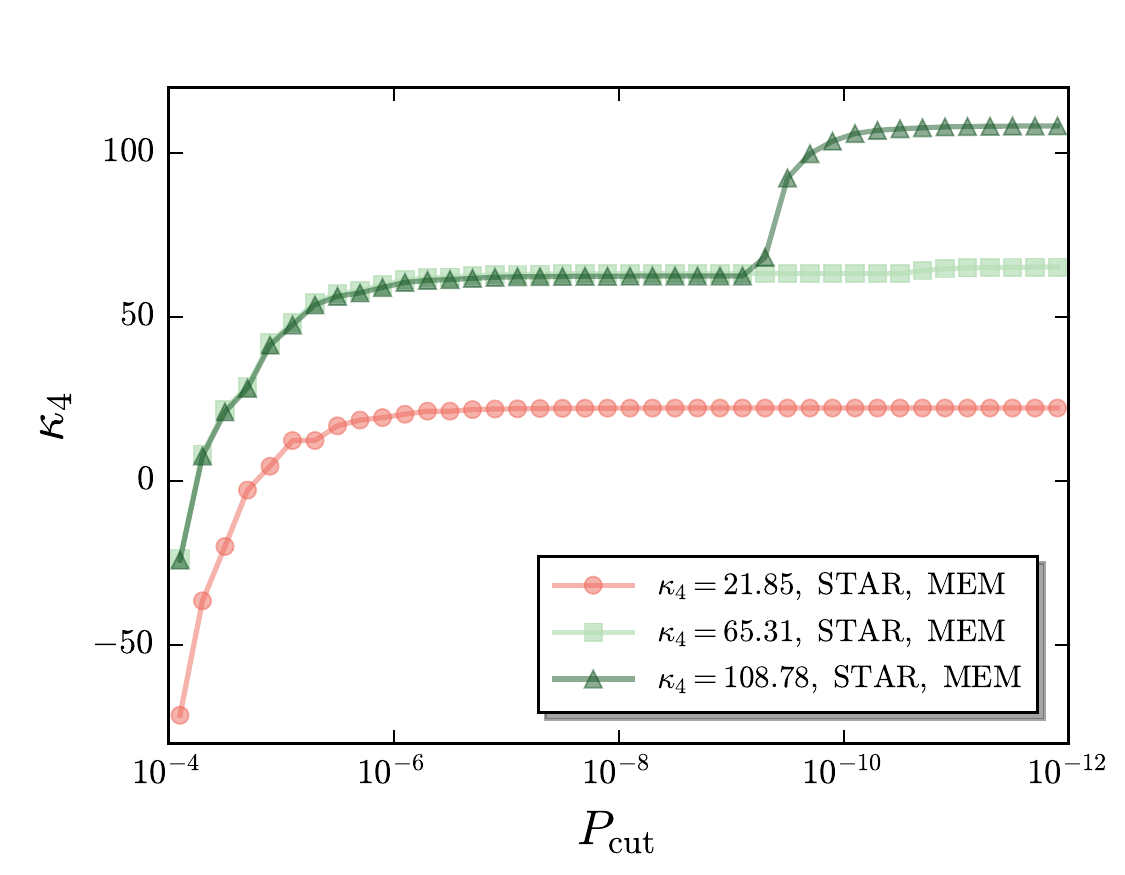}
\caption{Fourth-order cumulant as a function of the cut $P_{\mathrm{cut}}$ in \Eq{eq:Pcut}, where the baryon number distributions of the STAR data with three values of $\kappa_4$, shown in the left panel of \Fig{fig:rho-com}, are used. The MEM is employed to reconstruct the distributions.}\label{fig:Pcut-kappa4}
\end{figure}
%

In this section, we would like to apply the two methods, the MEM and GP, to experimental data, in order to reconstruct the distributions. Here we choose the most important and intriguing fluctuation measurements at the collision energy $\sqrt{s_{NN}}=7.7$ GeV in the Beam Energy Scan experiment at RHIC, which have been collected in \Tab{tab:fRG&Exp-kappa}. Here, the first four order cumulants of the net-proton distributions for central (0-5\%) Au+Au collisions are presented, see \cite{STAR:2020tga} for more details. The cumulants measured at $\sqrt{s_{NN}}=7.7$ GeV are believed to play a crucial role in the non-monotonic dependence of the kurtosis on the collision energy observed in experiments. Moreover, one can see that the error for the fourth-order cumulant of experimental data is very large. So it is very interesting to investigate the baryon number distributions reconstructed from the measured experimental data, especially the influence of the fourth-order cumulant. Note that here we do not distinguish the net-proton number and the net-baron number, since if the net proton is an ideal proxy for the net baryon in the experimental measurement, the difference between them is just a constant factor. In \Tab{tab:fRG&Exp-kappa} we also show the theoretical results at $\sqrt{s_{NN}}=7.7$ GeV obtained in the fRG \cite{Fu:2021oaw}.

In the left panel of \Fig{fig:rho-com} we show the baryon number distributions reconstructed with the MEM and based on the STAR cumulants in \Tab{tab:fRG&Exp-kappa}. Since the errors of the first three order cumulants of the STAR data are small, we just use their central values. By contrast, the error of the measured $\kappa_4$ is sizable, and it is more reasonable to choose several different values for $\kappa_4$, that scatter around its central value but within the error. It is found that with the increase of $\kappa_4$, there is another peak in the distribution function developed in the region of large $N_B$, the critical value of $\kappa_4$ for this behavior, with the values of $\kappa_1$ through $\kappa_3$ shown in \Tab{tab:fRG&Exp-kappa}, is about 60, close to the central value measured in experiments. This weird behavior of distribution function is reflected in the dependence of cumulants on the cut defined in \Eq{eq:Pcut} as well. In \Fig{fig:Pcut-kappa4} the fourth-order cumulant is depicted as a function of the cut. Apparently, it is found that when the value of $\kappa_4$ is large and above the critical value, the cumulant jumps up to another value after it has already saturated at $P_{\mathrm{cut}}\sim10^{-6}$, since the second peak in the distribution function begins to play a role.

We also use the GP to reconstruct the distribution function based on the STAR data. Unfortunately, reliable calculations in the region of $P(N_B)\lesssim 10^{-10}$ are still not accessible for the moment in the GP, where a boundary condition is implemented as discussed above, and thus one is not able to investigate the second peak in the GP. However, we also show the reconstructed distribution function of the STAR data within the GP in the right panel of \Fig{fig:rho-com}. It is observed that it is not consistent with the MEM result, and one can see that there is a kink in the curve of GP for the STAR data when $N_B \in (50,70)$. On the contrary, for the fRG cumulants in \Tab{tab:fRG&Exp-kappa}, both the GP and MEM produce almost identical baryon number distributions as shown in the right panel of \Fig{fig:rho-com}, and there is no second peak for the MEM reconstruction as shown in the left panel of \Fig{fig:rho-com}.

What can we learn from the results of the reconstructed baryon number distributions based on the experimental data above? Certainly, the behavior of a baryon number distribution with two or multiple peaks is odd or anomalous. Given some general assumptions, such as the maximum entropy principle, if a baryon number distribution has a weird behavior, e.g., multiple peaks, one may call this unnaturalness, which in turn can be used to constrain the cumulants of net-baryon or net-proton number distributions measured in experiments.

\section{conclusions and summary}
\label{sec:conclu}

In this work, we have employed two methods, the maximum entropy method and the Gaussian process regression which are both well-suited for treatment of inverse problems, to reconstruct net-baryon number distributions based on a finite number of cumulants of the distribution. 

Due to the property of convex optimization, one is able to obtain an analytic function of baryon number distribution within the approach of MEM. In the GP, the hyperparameters in the kernel function are determined by minimizing the negative logarithm likelihood function, which allow us to extract both the central value and the variance of $P(N_B)$ for each value of $N_B$.

We use the MEM and GP to reconstruct baryon number distributions across the chiral phase transition by employing the cumulants calculated in the fRG. It is found that the two different approaches, MEM and GP, produce almost identical distribution functions for the fRG inputs. The dependence of cumulants on the cut of distribution function $P_{\mathrm{cut}}$, i.e., the minimum value of $P(N_B)$ taken into account in the computation of cumulants, is investigated. We observe that the saturation $P_{\mathrm{cut}}$ decreases significantly with the increase of the order of cumulants. This implies that in the experimental measurements of high-order cumulants of the net proton number and other conserved charges, the long tails of the distribution would play a crucial role. Higher order is concerned, longer tail should be taken with care.

Finally, we apply the  MEM and GP to the experimentally measured cumulants at the collision energy $\sqrt{s_{NN}}=7.7$ GeV. Values of $\kappa_1$ through $\kappa_3$ are fixed at their central values, and the value of  $\kappa_4$ is varied around its central value and within its error. The calculation of MEM shows that with the increase of $\kappa_4$, there is another peak in the distribution function developed in the region of large $N_B$. Moreover, the reconstructed distribution function of the STAR data within the GP is not consistent with the MEM result. This unnaturalness observed in the reconstructed distribution function might be used to constrain the cumulants measured in experiments.


\begin{acknowledgments}
We thank Xiaofeng Luo for discussions. This work is supported by the National Natural Science Foundation of China under Grant No. 12175030. 
\end{acknowledgments}

	
\bibliography{ref-lib}

\end{document}